\documentclass[12pt]{iopart}
\usepackage{iopams}
\usepackage{fancybox}
\usepackage{eepic}
\usepackage{times}
\usepackage{latexsym}
\usepackage{pifont}
\usepackage{citesort}
\usepackage{pstricks}
\usepackage{epsf}

\newcommand{\bv}[1]{\mbox{\boldmath $#1$}}

\def\jpcm{J. Phys.: Cond. Matt.}   

\def\prb{Phys. Rev. B}

\def\prl{Phys. Rev. Lett.}

\def\jcp{J. Chem. Phys.}        
\newcommand{\bvsub}[1]{\mbox{\scriptsize\boldmath $#1$}}

\begin{document}
\article{}{An ignition key for atomic-scale engines}
\author{Daniel Dundas, Brian Cunningham, Claire Buchanan, Asako Terasawa,
Anthony T Paxton, Tchavdar N Todorov}
\address{Atomistic Simulation Centre, School of Mathematics and Physics, 
        Queen's University Belfast, 
        Belfast BT7 1NN, UK} 
\ead{d.dundas@qub.ac.uk}

\begin{abstract}
A current-carrying resonant nanoscale device, simulated by non-adiabatic 
molecular dynamics, exhibits sharp activation of non-conservative 
current-induced forces with bias. The result, above the critical bias,
is generalized rotational atomic motion with a large gain in kinetic energy. 
The activation exploits 
sharp features in the electronic structure, and constitutes, in effect, an
ignition key for atomic-scale motors. A controlling factor for the effect
is the non-equilibrium dynamical response matrix for small-amplitude atomic motion 
under current. This matrix can be found from the steady-state electronic structure
by a simpler static calculation, providing a way to detect the likely appearance, or 
otherwise, of non-conservative dynamics, in advance of real-time modelling.
\end{abstract}

Nanoscale conductors~\cite{agrait:2003,galperin:2007} carry current densities 
orders of magnitude larger than in macroscopic wires, resulting in substantial forces. 
The current-induced force on a nucleus consists of {\bfseries (I)} 
the average force, and {\bfseries (II)} force noise originating from the corpuscular nature of electrons~\cite{lu:2010,lu:2012}. 
{\bfseries II} causes inelastic electron-phonon scattering and Joule heating~\cite{agrait:2002,galperin:2007}. 
{\bfseries I} contains the so-called electron-wind force, and velocity-dependent 
forces~\cite{dundas:2009,lu:2010,todorov:2010,todorov:2011,bode:2011,bode:2012,lu:2011,lu:2012}. 
The wind force results from momentum transfer in elastic electron-nuclear scattering,
and drives electromigration~\cite{sorbello:1997}. Wind forces can be calculated 
from first principles, to study their effect on nanoscale devices~\cite{todorov:2001,diventra:2002,brandbyge:2003,sanvito:2011}. 

The electron-wind force is receiving fresh attention due to a remarkable property: 
it is non-conservative (NC) and can do net work on atoms around closed 
paths~\cite{sorbello:1997,dundas:2009,lu:2010,todorov:2010,todorov:2011,bode:2011,bode:2012,lu:2011,lu:2012}.
This phenomenon, which we call the waterwheel effect, opens up interesting questions. 
It provides a mechanism for driving
molecular engines~\cite{seideman:2003,kral:2005,bailey:2008,cisek:2011,vanderzant:2010,tierney:2011}. 
But the gain in kinetic energy of the atoms from the work done by NC forces is also a 
potential failure mechanism, possibly more potent than Joule heating. 
There are experimental indications of anomalous bias-activated apparent heating in point 
contacts~\cite{tsutsui:2008,tsutsui:2007}, above that expected from Joule heating 
alone~\cite{todorov:2001,smit:1998,smit:2004,sakai:2006}. The waterwheel effect is
a possible activation mechanism also for the electromigration phenomena that become
a central issue under large currents~\cite{taychatanapat:2007,kizuka:2009}.

The applied bias is a key factor for the operation
of NC forces~\cite{dundas:2009,lu:2010,todorov:2011,bode:2011,bode:2012,lu:2011,lu:2012}. 
First, they have to compete against the electronic friction (a velocity-dependent force), 
and this may require a critical current. Second, the waterwheel effect requires
pairs of normal modes degenerate in frequency~\cite{dundas:2009,lu:2010,todorov:2011}. 
If there is a frequency mismatch, a critical bias may be needed to overcome it. 
Ramping up the bias to overcome these factors is, notionally, like having to press the 
accelerator harder to climb a hill. 

The bias is a sensitive control parameter for the current-driven 
excitation of atomic motion in resonant systems~\cite{jorn:2009,jorn:2010}. 
In this letter we show how resonances can be exploited to turn the NC force on and off, 
akin to turning the engine of a car on and off. We will see that this ``switch'' is 
robust against decreasing resonance width. We will illustrate further how NC 
forces can be gauged to extract work from the current with no net angular momentum 
transfer to the real-space atomic motion. 

\begin{figure}[b]
\centerline{\epsfxsize=7cm\epsffile{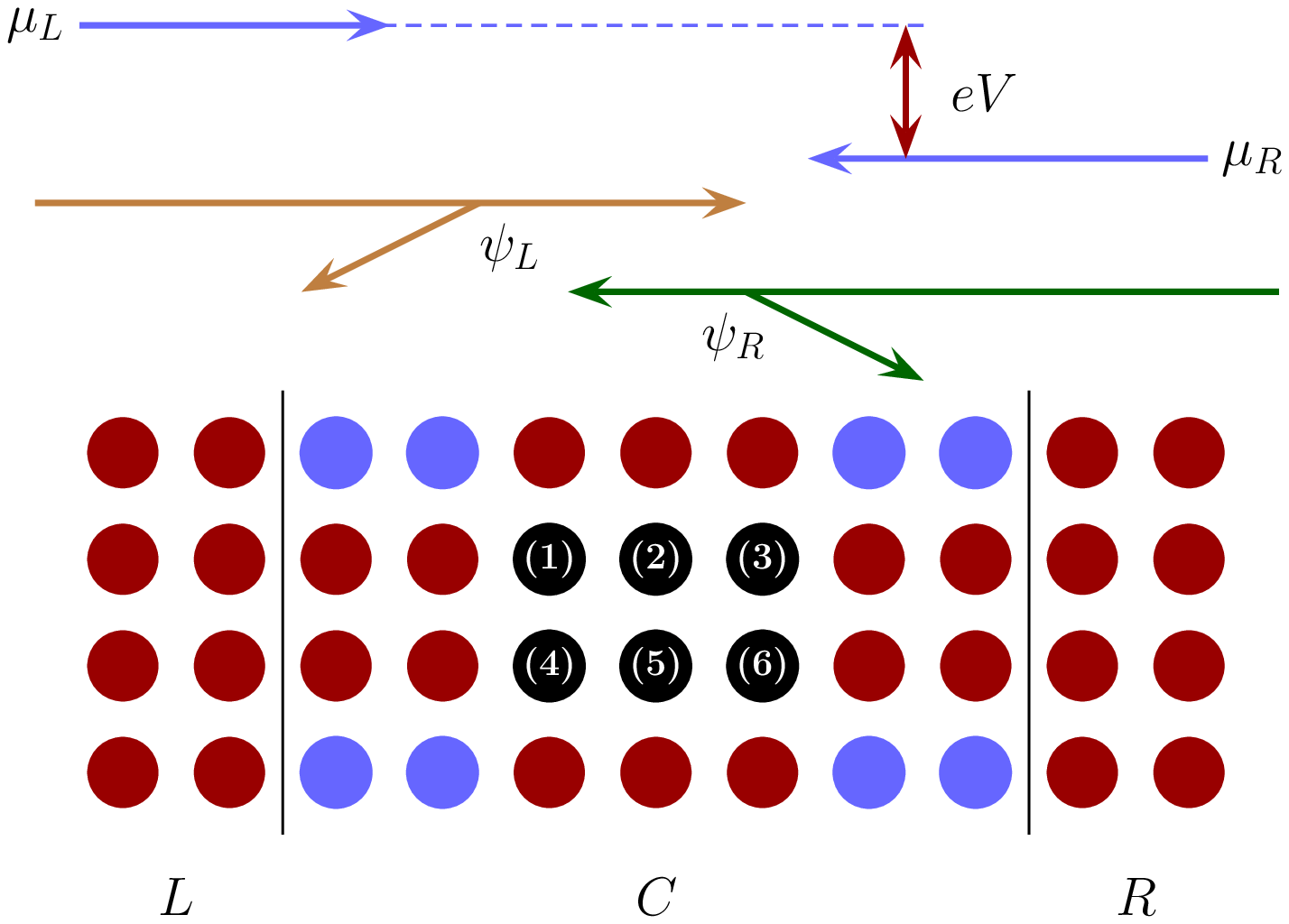}}
\caption{The 2D atomic strip considered in our calculations. 
A device $C$ is connected to two semi-infinite electrodes $L$ and $R$. Red denotes metallic 
regions, the blue atoms are insulating, and black denotes the atoms that will be treated dynamically in 
the simulations. In the Landauer picture, particle reservoirs inject electrons from the left and 
from the right with electrochemical potentials $\mu_{L}=\mu+eV/\mbox{2}$ and $\mu_{R}=\mu-eV/\mbox{2}$ 
respectively, where $\mu$ is the equilibrium Fermi level, $e (<\mbox{0})$ is 
the electron charge and $V$ is the applied bias. $eV > \mbox{0}$ corresponds to
electron flow from left to right. Electrons in the steady state are described by Lippmann-Schwinger
scattering states, $\{\psi_{L(R)}\}$, populated with the Fermi-Dirac distributions 
$f_{L(R)}$ corresponding to $\mu_{L(R)}$.
}
\label{fig:fig-qdot}
\end{figure}

Our system is shown in Fig.~\ref{fig:fig-qdot}. It is a 2D metallic nanowire
in the $x$-$y$ plane. The wire has a simple square lattice structure with a bondlength 
of 2.6 \AA. Electrons are described in a nearest-neighbour single-orbital tight-binding 
model with parameters for gold~\cite{sutton:2001}, except the band filling which here is 
set to 0.16. The hopping integral is $H = -$3.3175 eV and its derivative 
with distance is $H^\prime =$ 5.1038 eV/\AA. Red denotes metallic atoms, with an onsite 
energy set to zero. Region $C$ is a resonant device created by the blue atoms, 
which have an elevated onsite energy, $E_{\rm b}$. These insulating atoms form a 
double constriction. Current is supplied by the leads $L$ and $R$.

First we examine the steady-state transport properties of the system, in the Landauer picture 
summarized in Fig.~\ref{fig:fig-qdot}. Under bias $V$, the steady-state 1-electron density matrix 
can be written as
\begin{equation}
\hat{\rho}(V, \bv{R}) = \int_{-\infty}^{+\infty} \, \hat{\rho}(E) \,dE \,,
\end{equation}
where $\hat{\rho}(E) = f_L(E) \hat{D}_L(E) + f_R(E) \hat{D}_R(E)$ and $\hat{D}_{L(R)}(E)$ are 
the density of states operators for the stationary scattering states $\{ \psi_{L(R)} \}$, 
and $\bv{R}$ denotes ionic coordinates. 
We consider non-interacting electrons throughout. Spin is subsumed in $\hat{D}_{L(R)}(E)$.
 
Forces exerted by electrons on ions are described within the Ehrenfest 
approximation~\cite{ehrenfest:1927}. In the steady state, this force is given by
\begin{equation}
\bv{F}(V, \bv{R}) = \mbox{Tr}\Bigr\{\hat{\bv{F}}(\bv{R})\hat{\rho}(V, \bv{R})\Bigr\}\,,
\label{fe}
\end{equation}
where $\hat{\bv{F}}(\bv{R}) = -\nabla_{\bvsub{R}}\hat{H}_e(\bv{R})$ is the force operator, 
defined in terms of the electronic Hamiltonian $\hat{H}_e(\bv{R})$. The curl of the force on an ion with position $(R_{x},R_{y},R_{z})$,
\begin{equation}
\label{eq:curl-define}
(\nabla \times {\bf F})_{z} = 
\frac{\partial F_y}{\partial R_x} - \frac{\partial F_x}{\partial R_y}\,,
\end{equation}
is given by~\cite{dundas:2009,todorov:2010}
\begin{equation}
(\nabla \times {\bf F})_{z} =  
4\pi\, \int_{-\infty}^{+\infty}\,\mbox{Im}\,\mbox{Tr}\Bigr\{\hat{F}_x \hat{D}(E)\hat{F}_y \hat{\rho}(E)\Bigr\}\,dE\,.
\label{eq:nc0}
\end{equation}
Here, ${\hat D}(E)={\hat D}_{L}(E) + {\hat D}_{R}(E) = [\hat{G}^-(E) -\hat{G}^+(E)]/2\pi\mathrm{i}$ is the total 
density of states operator, where $\hat{G}^\pm(E)$ are the retarded and advanced
Green operators. The curl comes solely from the non-equilibrium part of the density matrix. Thus,
at zero electronic temperature
\begin{eqnarray}
(\nabla \times {\bf F})_{z} & = 4\pi\,\int_\mu^{\mu_L}\,\mbox{Im}\,\mbox{ Tr}\left\{\hat{F}_x\hat{D}(E)\hat{F}_y\hat{D}_L(E)\right\}\,dE\nonumber\\
& - 4\pi\,\int_{\mu_R}^\mu \, \mbox{Im}\,\mbox{ Tr}\left\{\hat{F}_x\hat{D}(E)\hat{F}_y\hat{D}_R(E)\right\} \,dE\,.
\label{eq:ncneq}
\end{eqnarray}
To lowest order in the bias, this simplifies to 
\begin{equation}
(\nabla \times {\bf F})_{z} =  
4\pi\,\mbox{Im}\,\mbox{Tr}\Bigr\{\hat{F}_x \hat{D}(\mu)\hat{F}_y \Delta {\hat \rho}\Bigr\}\,, 
\label{eq:ncneq_sb}
\end{equation}
where $\Delta{\hat \rho} = eV[{\hat D}_{L}(\mu) - {\hat D}_{R}(\mu)]/2$. 
\begin{figure}[t]
\centerline{\epsfxsize=8cm\epsffile{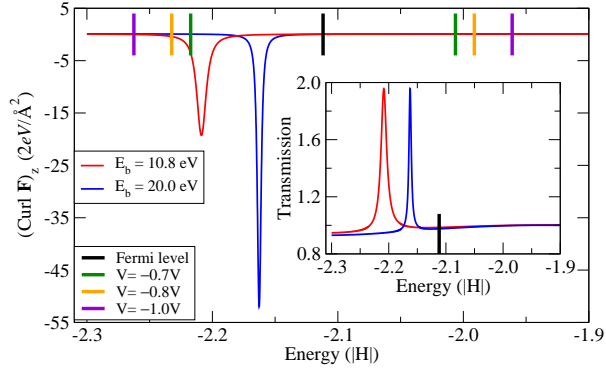}}
\caption{Curl of the current-induced force on atom (1) from
Fig.~\protect\ref{fig:fig-qdot} vs Fermi level, to lowest order in the bias $V$.
Under finite bias, the curl is given by the integral of the curve 
(with the integrand now in units of 2$/$\AA$^2$) from $\mu_{R}$ to $\mu_{L}$.
The inset shows the transmission function for the system. 
For the parameters in the text, the equilibrium Fermi level is 
denoted by the black line. The energy window for conduction under 
several biases (used in the simulations) is marked by coloured lines. 
The onsite energy shift on the barrier atoms is $E_{\rm b} = 10.8$ eV (red) 
 and $E_{\rm b} = 20$ eV (blue).}
\label{fig:fig-curl}
\end{figure}

Fig.~\ref{fig:fig-curl} shows the energy-resolved curl of the force on atom (1) 
from Fig.~\ref{fig:fig-qdot} and the transmission function for the system (inset), 
for two values of $E_{\rm b}$. 
By symmetry, the curl for atom (3) is identical. To see this, reverse the bias. 
Eq.~(\ref{eq:ncneq_sb}) changes sign. Atom (1) under the new bias is physically 
equivalent to atom (3) under the old bias. It follows that the curl on (3) and (1) 
is the same, under a given bias. The curl on (4) is the negative of that on (1), and so on.

At the given Fermi level, the perfect 4-atom wide strip has two open conduction 
channels, corresponding to the two lowest transverse modes in the wire. The double 
constriction formed by the insulating atoms constitutes a tunnelling double barrier for 
the higher mode. The resultant electronic resonance is present both in the curl and in 
the transmission in Fig.~\ref{fig:fig-curl}. Raising $E_{\rm b}$ makes the
walls harder and the effective constriction narrower, making the resonance narrower. 
The shift is the energy renormalization that accompanies 
changes in lifetime for quasibound states. (The effect eventually saturates with $E_{\rm b}$.)

First we study the system with the wider resonance, for $E_{\rm b} = 10.8$ eV. 
The peak in the curl is at an energy of $-$2.21$|H|$. Then we expect that
we can switch the NC force on and off by varying the bias so as to just include, 
or exclude, the peak from the conduction window. For the present system 
the equilibrium Fermi level is $\mu = -$2.1118$|H|$. The peak falls in the voltage 
range $-$1.0 V $< V < -$0.3 V. We thus expect a threshold bias 
somewhere in this range, at which the waterwheel effect is activated.

To test these ideas we simulate this system numerically. 
As in Ref.~\cite{dundas:2009}, the simulations are carried out in
the Ehrenfest approximation. Ehrenfest dynamics consists of the quantum Liouville 
equation for the electronic density matrix, coupled with Newtonian equations for nuclei. 
It includes the mean forces {\bfseries (I)}, 
but excludes the noise {\bfseries (II)}~\cite{horsfield:2005,lu:2012}.
The resultant suppression of Joule heating enables us to attribute any observed heating to 
work done by the NC forces. Current is generated by the multiple-probes open-boundary 
method~\cite{mceniry:2007},
with parameters: $S$ is a $4 \times 151$-atom strip; $C$ contains 28 atoms; the source and sink terms
are applied to the 248 end atoms in each electrode; $\Gamma = 0.6$ eV, $\Delta = 0.001$~eV.
\begin{figure}[t]
\centerline{\epsfxsize=8cm\epsffile{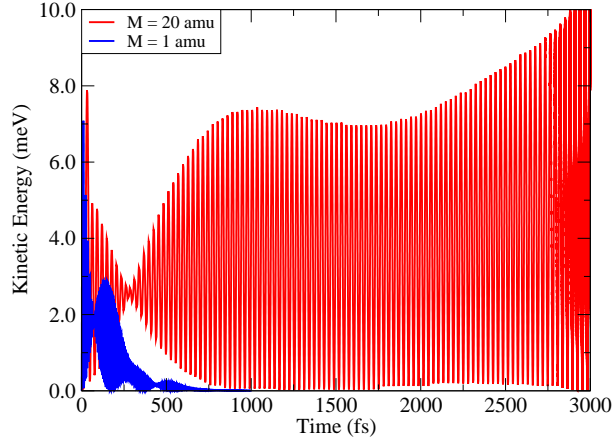}}
\caption{Kinetic energy of atom (1) from Fig.~\protect\ref{fig:fig-qdot} during current flow 
over a period of 3000 fs, under a bias of $-$1.0 V, for the wide resonance ($E_{\rm b} =$ 10.8 eV). In this simulation only atom (1) is allowed to move. 
The energy window for conduction encloses the peak in the curl of the 
current-induced force in Fig.~\protect\ref{fig:fig-curl} but when the mass of the atom is 
$M=1$ amu (blue) non-conservative heating does not occur. However, when $M= 20$ amu (red) 
the non-conservative force overcomes the electronic friction and the waterwheel effect is seen.}
\label{fig:fig-activate1}
\end{figure}

First we allow only atom (1) from Fig.~\ref{fig:fig-qdot} to move, setting its mass to 
$M = 1$ atomic mass unit (amu). Fig.~\ref{fig:fig-activate1} shows its kinetic energy in time 
for a bias of $-$1.0 V. Although the bias window encompasses 
the peak in the curl of the current-induced force, no heating occurs. Instead cooling 
under the electronic friction is observed. Velocity-dependent forces can be calculated
perturbatively under steady-state conditions~\cite{lu:2010,bode:2011,bode:2012,lu:2011,lu:2012}. 
Another way of isolating their effect is via the atomic mass~\cite{dundas:2009}. 
Increasing $M$ suppresses them relative to the velocity-independent NC force, 
enabling the latter to dominate. This is illustrated in Fig.~\ref{fig:fig-activate1} for $M=20$ amu.

\begin{figure}[t]
\centerline{\epsfxsize=8cm\epsffile{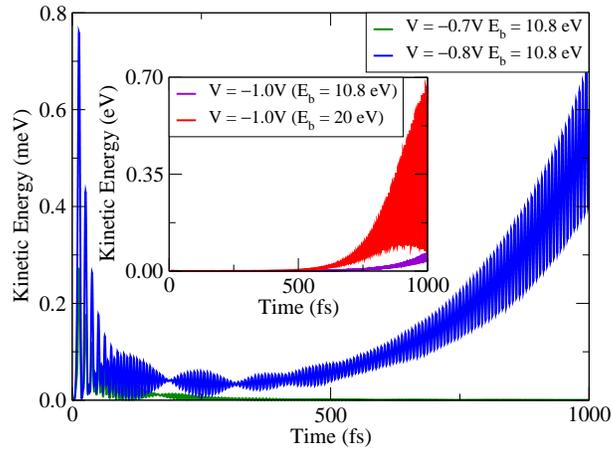}}
\caption{Total kinetic energy of atoms (1) and (3) in Fig.~\protect\ref{fig:fig-qdot}
for the wide resonance ($E_{\rm b} =$ 10.8 eV) during current flow over 1000 fs, for three applied 
biases: $-$0.7~V, $-$0.8~V (main panel) and $-$1.0~V (inset). 
When the magnitude of the bias is below 0.8~V we see only cooling, 
whereas for larger biases the heating due to non-conservative forces dominates. 
The inset shows the kinetic energy for the two atoms at $-$1.0 V 
for both the wide resonance and the narrow resonance ($E_{\rm b} =$ 20 eV) and we see that 
the kinetic energy gain for the narrow resonance exceeds that for the wide resonance. $M =$ 1 amu.}
\label{fig:fig-activate2}
\end{figure}

Next we allow atoms (1) and (3) to move, with $M = 1$ amu.  
Fig.~\ref{fig:fig-activate2} presents the total kinetic energy of the two atoms 
for biases of $-$0.7 V, $-$0.8 V and $-$1.0 V. For $-$0.7 V we see only cooling. 
At $-$0.8 V the kinetic energy starts to grow, showing that the waterwheel effect 
has been activated. Increasing the bias to $-$1.0 V shows an even larger energy gain. 
If we examine the trajectories we observe unidirectional rotational orbits that 
spiral outward as the energy grows. These are similar to those in bent atomic 
chains~\cite{dundas:2009}, although here they tended to be elliptical.

Thus for the two moving atoms we see the expected bias threshold for the waterwheel 
effect, between $-$0.7 V and $-$0.8 V. But for the same mass, for one moving atom 
the effect fails to manifest itself even above this threshold. 
Both features can be understood in terms of the non-equilibrium dynamical response matrix. 

We write the dynamical response matrix under current as
\begin{equation}
K_{\nu\nu'}(V, \bv{R}) = - \frac{\partial F_{\nu}}{\partial R_{\nu'}} + 
                           \frac{\partial^2 V_{\rm II}}{\partial R_{\nu}\partial R_{\nu'}} \,,
\label{eq:dyn-resp1}
\end{equation}
where $F_{\nu}$ is the electronic force of Eq.~(\ref{fe}), $V_{II}$ is the ion-ion interaction,
and index $\nu$ labels ionic degrees of freedom. 
$K_{\nu\nu'}$ may be split into an equilibrium part
$K_{\nu\nu'}^{\mbox{\scriptsize eq}}$ and a non-equilibrium correction $\Delta K_{\nu\nu'}$,
$K_{\nu\nu'}(V, \bv{R}) = K_{\nu\nu'}^{\mbox{\scriptsize eq}} + \Delta K_{\nu\nu'}$.
Since $\nabla \times {\bf F}$ is non-zero under current, Eq.~(\ref{eq:curl-define}) implies 
that $\Delta K_{\nu\nu'}$ is in general not symmetric and hence mode frequencies under current 
will in general be complex. Complex frequencies imply motion that grows or decays in time. 
The growing modes are the waterwheel, or runaway~\cite{lu:2010}, modes. 

\begin{figure}[b]
\centerline{\epsfxsize=8cm\epsffile{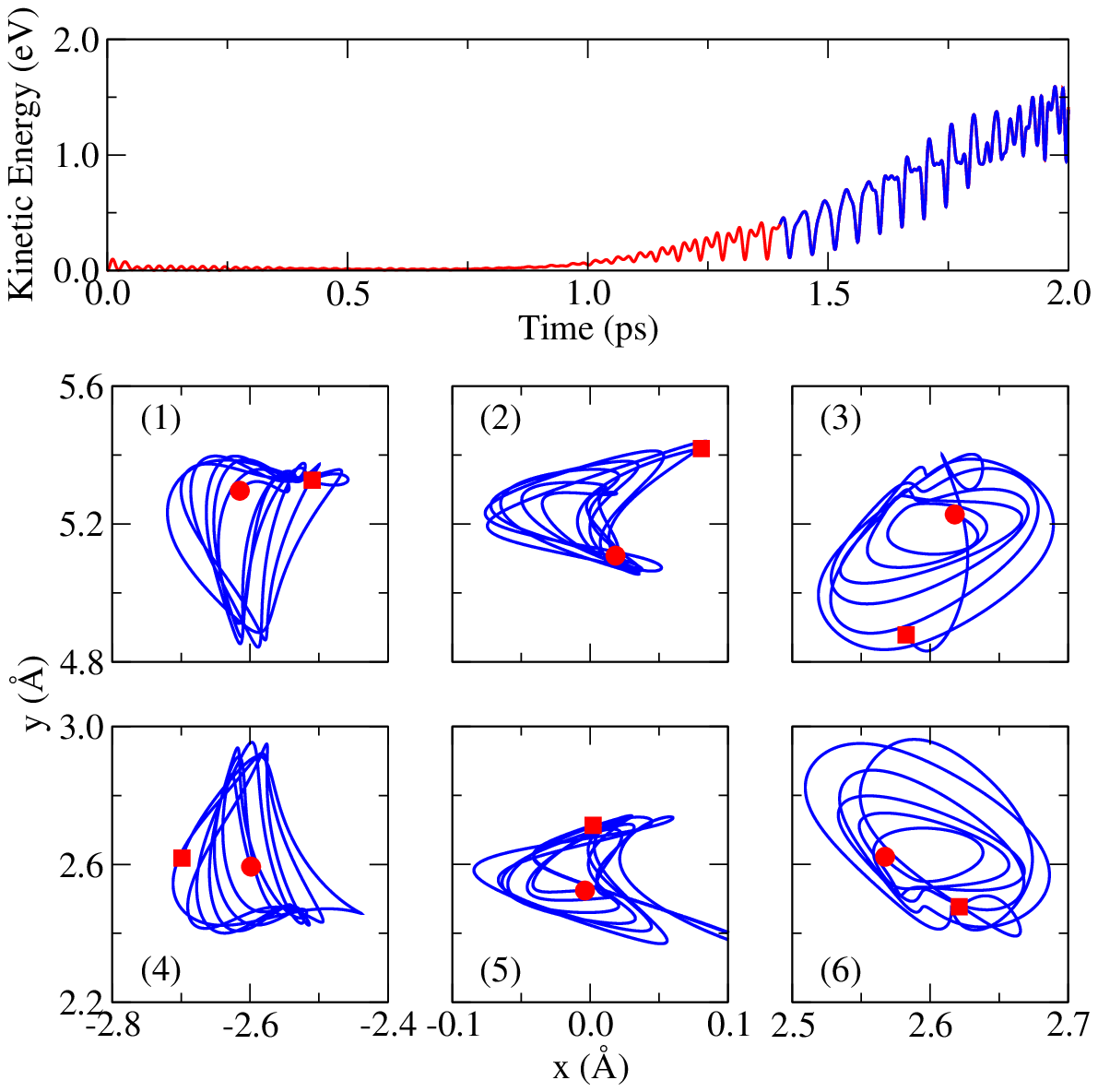}}
\caption{Total kinetic energy (top) and individual trajectories (bottom), 
with atoms (1)--(6) allowed to move,
for the wide resonance ($E_{\rm b} =$~10.8~eV). The bias is $-$1.0 V and $M =$ 20 amu.
The trajectories cover an advanced stage of the 
motion, with large displacements: this interval is denoted in blue in the kinetic energy plot. 
Dots (squares) denote the start (end) of each trajectory.}
\label{fig:fig-sixmove}
\end{figure}
Differentiating the steady-state density matrix by scattering theory as 
in~\cite{dundas:2009,todorov:2010}, we obtain
$\Delta K_{\nu\nu'} = S_{\nu\nu'} + A_{\nu\nu'}$, where the symmetric and antisymmetric 
components are given by 
\begin{eqnarray}
S_{\nu\nu'} &=& \displaystyle\frac{\Delta K_{\nu\nu'} + \Delta K_{\nu'\nu}}{2} \nonumber \\[0.2cm]
& = & \sum_{i=L,R}\int_\mu^{\mu_i} \, \mbox{ Tr}\left\{\hat{K}_{\nu\nu'}\hat{D}_i(E)\right\} \, dE \nonumber \\[0.2cm]
& + & 2 \sum_{i = L,R}\int_\mu^{\mu_i}\,\mbox{ Re Tr}\left\{\hat{F}_\nu\hat{R}(E)\hat{F}_{\nu'}\hat{D}_i(E)\right\} \, dE\label{eq:neq-sym}\\[0.4cm]
A_{\nu\nu'} &=& \displaystyle\frac{\Delta K_{\nu\nu'} - \Delta K_{\nu'\nu}}{2} \nonumber \\[0.2cm]
& = & 2\pi \sum_{i = L,R} \int_\mu^{\mu_i}\mbox{ Im Tr}\left\{\hat{F}_\nu\hat{D}(E)\hat{F}_{\nu'}\hat{D}_i(E)\right\} \, dE \label{eq:neq-asym}\,.
\end{eqnarray}
Above, $\hat{R}(E) = [\hat{G}^-(E) + \hat{G}^+(E)]/2$ and 
$\hat{K}_{\nu\nu'} = \partial^2 \hat{H}_{e}/\partial R_{\nu}\partial R_{\nu'} = -\partial \hat{F}_\nu/\partial R_{\nu'}$. 

The competition between $S$ and $A$ will 
determine whether mode frequencies are real or complex. We have calculated these frequencies 
for the systems above. When only atom (1) in Fig.~\ref{fig:fig-qdot} is allowed to move the equilibrium 
frequencies for $M=$ 1 amu are $\omega_1 = 0.456$ fs$^{-1}$ and $\omega_2 = 0.477$~fs$^{-1}$, 
while for a bias of $-$1.0 V $\omega_{1,2} = 0.480 \pm 0.005$i~fs$^{-1}$.

\begin{table*}[t]
\begin{tabular*}{0.95\textwidth}{@{\extracolsep{\fill}} ccccccc}
\hline\hline
Normal & \multicolumn{6}{c}{Applied Bias} \\[0.1cm]
  frequency                 &  \multicolumn{2}{c}{Equilibrium}  & $-$0.7V & $-$0.8V  & \multicolumn{2}{c}{$-$1.0V}\\[0.1cm]
 (fs$^{-1}$)                 &  W.R. & N.R. & W.R. & W.R. & W.R. & N.R. \\[0.1cm]\hline
$\omega_1$       & 0.442   & 0.432   &  0.368   & $0.470 + 0.016$i   &  $0.477 + 0.014$i   &  $0.476 + 0.013$i   \\
$\omega_2$       & 0.457   & 0.450   &  0.464   & $0.470 - 0.016$i   &  $0.477 - 0.014$i   &  $0.476 - 0.013$i   \\
$\omega_3$       & 0.473   & 0.471   &  0.474   & 0.482              &  0.483              &  0.481\\
$\omega_4$       & 0.494   & 0.492   &  0.481   & 0.490              &  0.500              &  0.504\\
\hline\hline
\end{tabular*}
\caption{Mode frequencies at equilibrium and at different applied biases when both atoms (1) and (3) 
in Fig.~\protect\ref{fig:fig-qdot} are allowed to move. Results are shown for both the wide resonance 
(labelled W.R.; $E_{\rm b} =$~10.8 eV) and for the narrow resonance (labelled
N.R.; $E_{\rm b} =$~20 eV). $M$ is 1 amu.}
\label{tab:tab-twomove}
\end{table*}
Table~\ref{tab:tab-twomove} shows the mode frequencies
when both atoms (1) and (3) are allowed to move. All frequencies are 
real at $-$0.7~V, while two frequencies become complex at $-$0.8~V. 
The waterwheel effect should activate in between, as observed in Fig.~\ref{fig:fig-activate2}.
The imaginary parts in Table~\ref{tab:tab-twomove} at $-$1.0 V are 
considerably larger than for one moving atom, explaining the greater 
ease with which waterwheel motion occurs.
Remarkably, the frequencies for the narrow and wide resonance in Table~\ref{tab:tab-twomove}
have comparable imaginary parts. We can see this from Fig.~\ref{fig:fig-curl}: 
the peak in the transmission is bounded, but the curl grows in height with decreasing width, 
keeping the area under the curve roughly constant. The reason is evident from 
Eq.~(\ref{eq:ncneq_sb}): the imaginary part comes from $\Delta {\hat \rho}$ 
and is related to the current; but present also is 
the density of states, which is not bounded. 
We conclude that, at least within limits, the ``switch'' 
for the NC force can be made more abrupt by making the resonance sharper.

The electronic friction should drop as the edges of the bias window 
move away from the resonance~\cite{mceniry:2008}. This is reflected
in the results for the two moving atoms in Fig.~\ref{fig:fig-activate2}. The imaginary 
parts of the frequencies at $-$0.8 V and $-$1.0 V (for both the wide and narrow resonance) 
are comparable. But the rate of energy gain increases as $\mu_L$ and $\mu_R$ retreat from the peak.

In Table~\ref{tab:tab-twomove} the large variation in 
$\omega_1$ around $-$0.7 V arises from the behaviour of the integrals 
in Eqs.~(\ref{eq:neq-sym}) and~(\ref{eq:neq-asym}), as the resonance moves into 
the conduction window. The interplay between the components of the dynamical response 
matrix will be studied further in future work. 

Allowing atoms (1), (3), (4) and (6) to move produces only real frequencies at $-$1.0 V,
but with all six atoms allowed to move two waterwheel modes appear, one with a
larger imaginary part than any above.
We simulate the system, starting with small random velocities, in
Fig.~\ref{fig:fig-sixmove}. To slow down the motion, we set $M=$ 20 amu.
The trajectories show that there is angular momentum transfer
to each atom, but no net angular momentum transfer overall.
Thus, although the elemental motion under NC forces is a generalized 
rotation in the abstract space of two normal coordinates coupled 
by current~\cite{todorov:2011}, these forces can do work without 
angular momentum transfer to the real-space dynamics of the atomic subsystem as a whole.

We have simulated dynamically a resonant device and have shown 
how the bias can switch on and off the non-conservative forces on atoms 
induced by current. This switch appears robust with decreasing resonance width.
This robustness will ultimately be limited by electron-phonon and
electron-electron interactions, which will modify the resonant energies 
and the resonance widths. An assessment of these factors is an important
direction for further work.
The abrupt activation of the waterwheel effect is a strong candidate mechanism 
behind anomalous heating in point contacts~\cite{tsutsui:2008,tsutsui:2007},
and can furnish a bias-controlled ``switch'' 
for current-driven atomic-scale motors. 

The non-equilibrium dynamical response 
matrix is a useful simple probe into non-conservative effects, and is a
generalization of the usual description of harmonic motion 
at equilibrium. A complete analysis of the eigenmodes under current
requires the inclusion of velocity-dependent forces, which
can be done perturbatively in the steady state~\cite{lu:2010,bode:2011,bode:2012,lu:2011,lu:2012}.
The simpler picture obtained by neglecting the 
velocity-dependent forces then amounts to taking the adiabatic limit of large atomic mass, 
or small atomic velocities. 
The eigenmode analysis now does not contain, for example, the electronic friction present
in our non-adiabatic molecular dynamics simulations, but captures rigorously and exactly the
physics of steady-state conduction in the Born-Oppenheimer limit. \\

We are grateful for support from the Engineering and Physical Sciences Research Council, under grant EP/I00713X/1.
It is a pleasure to thank Jingtao L\"{u}, Mads Brandbyge and Per Hedeg{\aa}rd for invaluable discussions.

\end{document}